\documentclass[preprintnumbers,prd,floatfix,nofootinbib,superscriptaddress]{revtex4}

\usepackage[paperwidth=210mm,paperheight=297mm,centering,hmargin=2cm,vmargin=2.5cm]{geometry}

\usepackage[tbtags]{amsmath}  
\usepackage{tabularx}
\usepackage{amssymb}          
\usepackage{bm}               
\usepackage{graphicx}         
\usepackage[export]{adjustbox}
\usepackage{hhline,multirow}  
\usepackage{dcolumn}          
\usepackage{slashed}
\usepackage{datetime}
\usepackage{color}
\usepackage[dvipsnames]{xcolor}
\usepackage{slashed}
\usepackage{subfloat}
\usepackage{subeqnarray}

\usepackage{amscd}            
\usepackage{epsfig}
\usepackage{dcolumn}          
\usepackage[dvipsnames]{xcolor}
\usepackage[normalem]{ulem}
\usepackage{mathtools,cases}
\usepackage{comment}
\usepackage[utf8]{inputenc}

\RequirePackage{color}
\RequirePackage[colorlinks=true
,urlcolor=black
,anchorcolor=black
,citecolor=black
,filecolor=black
,linkcolor=black
,menucolor=black
,pagecolor=black
,linktocpage=black
,pdfproducer=medialab
,pdfa=true
]{hyperref}

\newcommand{\timeord}{{\cal T}}
\newcommand{\antitimeord}{{\cal \overline{T}}}
\newcommand{\colav}{\frac{\text{Tr}_c}{N_c}}

\newcommand{\Tr}{{\rm Tr}}
\newcommand{\vect}[1]{\boldsymbol{#1}}

\newcommand{\sign}{\text{sign}}

\newcommand{\id}{{\mathbb{I}}}
\newcommand{\rom}{{|\Omega\rangle}}
\newcommand{\lom}{{\langle\Omega|}}

\newcommand{\psibar}{{\overline{\psi}}}
\renewcommand{\i}{{\mathrm i}}

\def\beq{\begin{equation}}
\def\eeq{\end{equation}}
\def\bea{\begin{eqnarray}}
\def\eea{\end{eqnarray}}
\def\nn{\nonumber}

\newcommand{\nocontentsline}[3]{}
\newcommand{\tocless}[2]{\bgroup\let\addcontentsline=\nocontentsline#1{#2}\egroup}

\allowdisplaybreaks[2]

\begin{document}


\preprint{JLAB-THY-23-3890}

\title{Gauge invariant spectral analysis of quark hadronization dynamics}

\author{Alberto Accardi}
\thanks{Electronic address: accardi@jlab.org - \href{https://orcid.org/0000-0002-2077-6557}{ORCID: 0000-0002-2077-6557}} 
\affiliation{Hampton University, Hampton, VA 23668, USA}
\affiliation{Theory Center, Thomas Jefferson National Accelerator Facility, 12000 Jefferson Avenue, Newport News, VA 23606, USA}

\author{Caroline S. R. Costa}
\thanks{Electronic address: costa@jlab.org - \href{https://orcid.org/0000-0003-1392-837X}{ORCID: 0000-0003-1392-837X}} 
\affiliation{Theory Center, Thomas Jefferson National Accelerator Facility, 12000 Jefferson Avenue, Newport News, VA 23606, USA}

\author{Andrea Signori}
\thanks{Electronic address: andrea.signori@unito.it - \href{https://orcid.org/0000-0001-6640-9659}{ORCID: 0000-0001-6640-9659}} 
\affiliation{Department of Physics, University of Turin, via Pietro Giuria 1, I-10125 Torino, Italy}
\affiliation{INFN, Section of Turin, via Pietro Giuria 1, I-10125 Torino, Italy}

\begin{abstract}
We study the Dirac decomposition of the gauge invariant quark propagator, whose imaginary part describes the hadronization of a quark as this interacts with the vacuum, and relate each of its coefficients to a specific sum rule for the chiral-odd and chiral-even quark spectral functions. 
Working at first in light-like axial gauge, we obtain a new sum rule for the spectral function associated to the gauge fixing vector, and show that its second moment is in fact equal to zero. Then, we demonstrate that the first moment of the chiral-odd quark spectral function is equal in any gauge to the so-called inclusive jet mass, 
which is related to the mass of the particles produced in the hadronization of a quark. 
Finally, we present a gauge-dependent formula that connects the second moment of the chiral-even quark spectral function to invariant mass generation and final state rescattering in the hadronization of a quark.
\end{abstract}

\date{\today}
\maketitle
\tableofcontents

\newpage
\section{Introduction}
\label{s:intro}

The confinement of the strong force directly connects the propagation of quarks and gluons with their hadronization, namely their transmutation into massive and colorless hadrons. 
On the one hand, the properties of partonic propagators in QCD can be theoretically investigated with techniques ranging from continuum methods~\cite{Alkofer:2000wg,Cloet:2013jya,Duarte:2022yur, Ferreira:2023fva, Horak:2022aza,Eichmann:2021zuv,Lessa:2022wqc,Fischer:2020xnb,Horak:2022myj,Rojas:2013tza} to effective theories and model calculations~\cite{Klevansky:1992qe,Muller:2010am,Tissier:2010ts,Siringo:2016jrc,Hayashi:2020few} and to lattice calculations~\cite{Skullerud:2002ge,Marques:2023cmi, Falcao:2020vyr,Oliveira:2021job,Bicudo:2015rma,Dudal:2019gvn,Duarte:2016iko,Oliveira:2012eh,Ayala:2012pb,Dudal:2010tf,Aguilar:2022thg}.
On the other hand, hard scattering processes with hadrons in the final state allow one, at least in principle, to probe the discontinuity of these propagators thanks to the optical theorem. In fact, as shown in Refs.~\cite{Accardi:2019luo,Accardi:2020iqn,Accardi:2017pmi}, 
the moments of the quark propagator's spectral functions can be explicitly connected to specific integrals of quark fragmentation functions (FFs) and furthermore directly enter the cross section of certain inclusive hard scattering processes. 
Thus, apart from their intrinsic interest, understanding the analytic properties of the quark propagator and of its associated spectral functions becomes of practical relevance for the phenomenology of hadron structure and of hadronization~\cite{Metz:2016swz,Accardi:2018gmh}.

In this paper, we elaborate on the gauge invariance of the so-called fully inclusive jet correlator introduced in Refs.~\cite{Accardi:2019luo,Accardi:2020iqn}, namely of 
\begin{align}
  \label{e:invariant_quark_correlator}
  \Xi_{ij}(k;w) = \text{Disc} \int \frac{d^4 \xi}{(2\pi)^4} e^{\i k \cdot \xi} \,
    \colav\, \lom 
    \big[\, \timeord\, W_1(\infty,\xi;w) \psi_i(\xi) \big]\, 
    \big[\, \antitimeord\, \psibar_j(0) W_2(0,\infty;w) \big] 
    \rom \ ,  
\end{align}
where $\rom$ is the interacting vacuum state of QCD, $\psi$ the quark field, $W_{1,2}$ are Wilson lines that ensure the gauge invariance of the correlator, and $w$ is an external vector that determines the direction of their paths, as discussed in detail later. $\timeord$ represents the time ordering operator for the fields whereas $\antitimeord$ represents the anti time ordering operator~\cite{Collins:2011zzd,Echevarria:2016scs}. For sake of brevity we omit the flavor index of the quark fields and of $\Xi$. The color trace averages over the incoming quark color quantum number.
One can show that the correlation function in Eq.~\eqref{e:invariant_quark_correlator} is related to the correlator for the single-hadron quark fragmentation functions at the operator level, by a sum over the flavor and spin of the produced hadrons and an on-shell integration over the four-momentum of the same hadrons~\cite{Accardi:2020iqn}. Thus, $\Xi_{ij}(k;w)$ represents mathematically the inclusive limit of the hadronization mechanism.
Moreover, as extensively discussed in Refs.~\cite{Accardi:2019luo,Accardi:2020iqn}, the jet correlator $\Xi$ can be written as the discontinuity of the gauge-invariant quark propagator:  
\begin{align}
  \Xi_{ij}(k;n_+) = \text{Disc} \int \frac{d^4 \xi}{(2\pi)^4} e^{\i k \cdot \xi} \,
  \colav\, \lom \psi_i(\xi) \psibar_j(0) W(0,\xi;n_+) \rom \, ,
\label{e:invariant_quark_correlator_W}
\end{align}
where the Wilson line is $W = W_2 W_1$.

The form of $\Xi$ given in Eq.~\eqref{e:invariant_quark_correlator} describes the hadronization of a quark into an unobserved jet of particles. It is relevant to calculate the cross section of processes with unidentified jets in the final states, for example for inclusive electron-proton Deep-Inelastic Scattering (DIS)~\cite{Collins:2007ph,Accardi:2008ne,Accardi:2017pmi,Accardi:2018gmh} and of Semi-Inclusive electron-positron Annihilation (SIA)~\cite{Accardi:2017pmi,USBelleIIGroup:2022qro,Accardi:2022oog}. Its emergence from factorization theorems is connected with the endpoint kinematics for the considered processes, see e.g. Refs.~\cite{Sterman:1986aj,Chay:2005rz}. 

The form of $\Xi$ given in Eq.~\eqref{e:invariant_quark_correlator_W}, instead, is a gauge-invariant generalization of the quark propagator, and makes possible the mentioned connection between quark propagation in the vacuum and hadronization. This connection is valid at the operator level and, after some formal manipulations, produces a sum rule that links the first moment of the chiral-odd quark spectral function (identified as the nonperturbative mass of the propagating quark) to the average of the produced hadron masses weighted by the chiral-odd scalar quark fragmentation functions $E$~\cite{Accardi:2019luo,Accardi:2020iqn}. 

While in Refs.~\cite{Accardi:2019luo,Accardi:2020iqn} the calculations were limited to the light cone gauge, in this paper the study of the gauge invariant propagator $\Xi$ is extended to a generic gauge. 
It is shown that the sum rule for the chiral-odd spectral function presented in Refs.~\cite{Accardi:2019luo,Accardi:2020iqn} is in fact formally valid in any gauge.
Moreover, a novel sum rule for the quark spectral function associated to the gauge fixing vector in light-like gauges is derived. 
Finally, a complete calculation of the twist-4 component of the jet correlator is presented, together with a sum rule for the second moment of the chiral-even spectral function.

This research line connects two aspects of QCD (and two research communities) which are intertwined, namely the study of the analytic properties of the quark and gluon propagators and the study of hadronization via scattering processes with hadrons in the final states. 
This article is devoted in particular to the properties of the quark propagator, and addresses in a sistematic way the calculation of the quantities which are relevant for hard scattering processes, namely the moments of the spectral functions and their gauge independence. 
The motivation behind this work lies, in part, in the possibility to express the jet mass as a moment of the chiral-odd quark spectral function in a generic gauge and not only in the light-cone gauge. In this gauge, in fact, computations are considerably more involved~\cite{Alkofer:2000wg,Mezrag:2023nkp} and, to the best of our knowledge, absent for quark spectral functions. Our results provide additional motivation to push forward with these calculations.  
Furthermore, the moments are also instrumental to determine mass corrections to semi-inclusive processes at sub-leading twist and to investigate higher-twist fragmentation functions~\cite{Mulders:1995dh,Metz:2016swz}. 
This connection is timely given the growing interest for higher-twist effects from the point of view of perturbation theory~\cite{Ebert:2021jhy,Gamberg:2022lju,Rodini:2023plb}, the current and future experimental measurements~\cite{Dudek:2012vr,Hadjidakis:2018ifr,HERMES:2020ifk,AbdulKhalek:2021gbh,Accardi:2022oog,USBelleIIGroup:2022qro,Burkert:2022hjz,Accardi:2023chb} and the recent  phenomenological analyses in this direction~\cite{Accardi:2017pmi,Bauer:2022mvl,Accardi:2022oog,USBelleIIGroup:2022qro}.

The paper is organized as follows.
In Sec.~\ref{s:jetcor} the Dirac structure of the gauge-invariant quark propagator is discussed, together with the related coefficient functions.  
Sec.~\ref{s:axial_gauges} features the spectral representation for the propagator, taking into account also the contribution from a light-like gauge-fixing vector $v$.
Sec.~\ref{s:generic_gauge} presents a comparison between the expressions for the coefficient functions obtained in the light-cone gauge~\cite{Accardi:2019luo,Accardi:2020iqn} with those obtained in a generic gauge. 
In Sec.~\ref{s:conclusions} the results are summarized.  

\section{Gauge invariant quark propagator}
\label{s:jetcor}

As shown in Refs.~\cite{Accardi:2019luo,Accardi:2020iqn}, the gauge invariant quark propagator \eqref{e:invariant_quark_correlator_W} can be further rewritten as a convolution of a quark bilinear operator and the Fourier transform of a Wilson line connecting the quark fields:
\begin{align}
   \Xi_{ij}(k;w) = \mathrm{Disc} \int d^4p  \colav
   \langle\Omega|i\widetilde{S}_{ij}(p;v)\widetilde{W}(k-p;w,v)
   |\Omega\rangle,
\label{def:jetcorr}
\end{align}
where
\begin{align}
    i\widetilde{S}_{ij}(p,v) & = \int\frac{d^{4}\xi}{(2\pi)^4}\,e^{i\xi\cdot p}\, \timeord\,\psi_{i}(\xi)\overline{\psi}_{j}(0),\\
    \widetilde{W}(k-p;\,w,v) & =  \int\frac{d^{4}\xi}{(2\pi)^4}\,e^{i\xi\cdot(k- p)}\,W(0,\xi;w,v).
\label{eq:w} 
\end{align}
In the definition~\eqref{def:jetcorr}, $k$ denotes the quark 4-momentum, $|\Omega\rangle$ is the interacting vacuum state, and the tilde sign marks functions in the momentum space. The 4-vector $w$ defines the direction of the Wilson line~\cite{Accardi:2019luo,Accardi:2020iqn}, which is introduced in order to guarantee the gauge invariance of the correlator, and will further be discussed below. One then easily sees that $\Xi$ provides one with a gauge-invariant definition of the two-point quark correlator $\langle \Omega | i\widetilde S | \Omega \rangle$. 
The quark operator $i\widetilde S$ and the Wilson line $\widetilde W$ may furthermore depend on the 4-vector $v$ defining an axial gauge; in non-axial gauges $v$ can still formally be used as a label reminding one of the dependence of these two quantities on the gauge fixing condition. 

The convolution representation in Eq.~\eqref{def:jetcorr} is convenient because it allows a direct connection between the gauge invariant quark correlator $\Xi$ and the spectral representation of the gauge-dependent quark propagator, that will be exploited and further studied below. The quark spectral representation has been extensively explored in recent years since its properties and analytical structure can possibly shed light on confinement~\cite{Horak:2022aza,Falcao:2022gxt,Solis:2019fzm,Sauli:2022vfy,Siringo:2016jrc,Sauli:2018gos,Hayashi:2020few}. Spectral properties of gauge invariant quark correlators have also been discussed in \cite{Yamagishi:1986bj,Sazdjian:2010ku} beside the aforementioned Refs.~\cite{Accardi:2019luo,Accardi:2020iqn}.  
It is also worth emphasizing once more that the jet correlator $\Xi$ is in itself is gauge invariant, whereas the quark operator $i\widetilde S$ is not, and therefore the LHS of Eq.~\eqref{def:jetcorr} is independent of $v$. Building on Refs.~\cite{Accardi:2019luo,Accardi:2020iqn}, we will exploit this fact to derive novel sum rules for the quark spectral functions in Sec.~\ref{s:axial_gauges} and~\ref{s:generic_gauge}.

Following Refs.~\cite{Accardi:2019luo,Accardi:2020iqn}, we work in light-cone coordinates (see Appendix~\ref{a:conv}) and boost the quark to large momentum in the light-cone minus direction so that its components satisfy $k^{-}\gg|\bm{k_T}|\gg  k^{+}$, where $|\bm{k_T}| \sim O(\Lambda)$ and $k^{+}, k^2\sim O(\Lambda^2)$, and $\Lambda$ is a power counting scale of order of the nonperturbative QCD scale $\Lambda_\text{QCD}$. We can then consider the gauge-invariant correlator integrated  over the subdominant $k^{+}$ component of the quark momentum~:
\begin{align}
    J_{ij}(k^{-},\bm{k_T};n_{+}) 
    &\equiv \frac{1}{2}\int dk^{+}\Xi_{ij}(k;w=n_{+}) ,
\label{eq:TMDcor}
\end{align}
which phenomenologically describes the inclusive hadronization of a high-energy quark into a jet of particles along the quark direction of motion, and we call ``inclusive jet correlator''.  It is precisely the integration over $k^+$ that allows one to derive sum rules for the quark spectral functions. 
Note that, in the definition of $J$, we follow Ref.~\cite{Accardi:2020iqn} and choose the Wilson line to lie in the plus light-cone $w=n_+$ direction. The full shape of the considered Wilson line is discussed in detail in the mentioned reference, but only its projection on the light-cone plus axis and the transverse plane matter in the calculations to be performed in this paper. Namely we will only need to consider the simpler $W_\text{TMD}(\xi^+,\bm{\xi_T}) \equiv W(0^-,\xi^+,\bm{\xi_T})$ transverse-position-dependent Wilson line and the $W_\text{coll}(\xi^+) \equiv W(0^-,\xi^+,\vect{0}_T)$ light-cone Wilson lines, defined as
\begin{align}
    W_\text{TMD}(\xi^+,\bm{\xi_T})         
        & = \mathcal{U}_{n_{+}}[0^-, 0^+, \vect{0}_T;0^-,\infty^+,\vect{0}_T] 
        \, \mathcal{U}_{\bm{n_T}} [0^-,\infty^+,\vect{0}_T;0^-,\infty^+,\bm{\xi_T}]
        \, \mathcal{U}_{n_{+}}[0^-,\infty^+,\bm{\xi_T};0^-,\xi^+, \bm{\xi_T}]
        \label{def:WTMD}\\
    W_\text{coll}(\xi^+) 
        & \equiv  W_\text{TMD}(\xi^+,\bm{0_T}) 
        = \mathcal{U}_{n_{+}}[0^-, 0^+, \vect{0}_T;0^-,\xi^+,\vect{0}_T] \ ,
        \label{def:Wcoll}
\end{align}
where $\mathcal{U}_v [a;\infty]$ is a straight Wilson line from $a$ to infinity along the $v$ direction,
\begin{align}
    \mathcal{U}_v [a;\infty] 
    = 
    \mathcal{P}\exp{\left(-ig\int_{0}^{\infty}ds\, v^{\mu}A_{\mu}\big(a+sv \big)\right)},
\end{align}
where $\mathcal{P}$ denotes the path-ordering operator, and with slight abuse of notation $\mathcal{U}_v [a;b] = \mathcal{U}_v [a;\infty] \, \mathcal{U}_v [\infty;b]$.
In Eqs.~\eqref{def:WTMD} and \eqref{def:Wcoll} one recognizes, respectively, the staple Wilson line used in TMD factorization theorems, and the light-cone Wilson line used in collinear factorization theorems. 

With these Wilson lines, the integrated correlator in Eq.~\eqref{eq:TMDcor} can then be used in perturbative calculations of inclusive DIS structure functions \cite{Accardi:2017pmi} and semi-inclusive electron-positron annihilation~\cite{Accardi:2022oog,USBelleIIGroup:2022qro}, where it coiuples respectively, to the nucleon's transversity parton distribution function and the polarized (e.g. $\Lambda$) hadron transversity fragmentation function. 
In these processes, the ``inclusive jet correlator'' $J$ 
is used instead of the free quark propagator to describe the hadronization of a scattered quark in the so-called end-point kinematics of the process~\cite{Sterman:1986aj,Chen:2006vd,Accardi:2008ne,Collins:2007ph}, where the invariant mass of the final state is limited, and the produced hadrons are kinematically constrained into a narrow -- yet unobserved -- jet of particles along the quark's direction of motion, thus earning its name.

Keeping in mind the outlined phenomenological applications -- but independently of these -- one can leverage the strong $k^{-} \gg |\bm{k_T}| \gg k^{+}$ ordering to give the inclusive jet correlator in Eq.~\eqref{eq:TMDcor} a ``twist'' decomposition controlled by a power counting scale $\Lambda$~\cite{Accardi:2019luo,Accardi:2020iqn}:
\begin{align}
    J(k^{-},\bm{k_T};n_{+}) 
    & = \dfrac{1}{2}\alpha(k^-)\gamma^+ + \dfrac{\Lambda}{k^-}\left[\zeta(k^-)\mathbb{I}+\alpha(k^-)
    \dfrac{\vect{\slashed{k}_T}}{\Lambda} \right]+\dfrac{\Lambda^2}{2(k^-)^2}\omega(k^-,\bm{k_T}^2)\gamma^- \ ,
\label{eq:jet_decomp}
\end{align}
where $\gamma^\mu$ are Dirac matrices and we suppressed the Dirac indices for simplicity.
Notice that the coefficients $\alpha$ and $\zeta$ are only functions of $k^-$, while $\omega$ is a function of both $k^{-}$ and $\bm{k_T}$. This is indeed the case as it will be further discussed below and will also be shown in Sec.~\ref{s:generic_gauge} by an explicit calculation of these coefficients.
Notice that we did not include time-reversal odd (T-odd) structures in the decomposition \eqref{eq:jet_decomp}, since these are not allowed in the fully inclusive hadronization of a quark described by the inclusive jet function $J$~\cite{Accardi:2020iqn}. On the contrary, T-odd structures would be allowed in less inclusive observables, such as with identified jet (see e.g. Refs.~\cite{Liu:2021ewb,Lai:2022aly}) or in one- and two-particle semi-inclusive measurements~\cite{Metz:2016swz}. Likewise we have not included any parity odd (P-odd) structures.

The $\alpha(k^-), \zeta(k^-)$ and $\omega(k^-,\bm{k_T}^2)$ are, respectively, the twist-$2$, twist-$3$ and twist-$4$ coefficients of the jet correlator, and can be obtained by projecting Eq.~(\ref{eq:TMDcor}) onto suitable Dirac structures. Denoting the projection of $J$ onto a generic Dirac matrix $\Gamma$ by
\bea
    J^{[\Gamma]} \equiv \Tr \left[J\, \frac{\Gamma}{2} \right] = \frac{1}{2}\int dk^{+}\Tr{\left[\Xi\, \frac{\Gamma}{2} \right]},
\label{eq:JGamma}
\eea
one finds:
\begin{align}
    \alpha(k^-) 
    &= J^{[\gamma^-]} = \frac{1}{2}\int dk^+\Tr{\left[\Xi\, \frac{\gamma^-}{2}\right]}\, ,
    \label{def:alpha}
    \\
    \zeta(k^-) 
    &= \frac{k^-}{\Lambda}J^{[\mathbb{I}]} = \frac{k^-}{\Lambda}\frac{1}{2}\int dk^+\Tr{\left[\Xi\, \frac{\mathbb{I}}{2}\right]}\, ,
    \label{def:zeta}
    \\
    \omega(k^-, \bm{k_T}^2) 
    &= \left(\frac{k^-}{\Lambda}\right)^2J^{[\gamma^+]} = \left(\frac{k^-}{\Lambda}\right)^2\frac{1}{2}\int dk^+\Tr{\left[\Xi\, \frac{\gamma^+}{2}\right]}\, .
\label{def:omega}
\end{align}
The specific dependence of $\alpha$ and $\zeta$ on $k^{-}$ and of $\omega$ on $k^-$ and $\bm{k_T}$ can be understood by considering the decomposition of the jet correlator $\Xi(k, n_{+})$ appearing in the integrands of Eqs.~\eqref{def:alpha}-\eqref{def:omega} on a basis of Dirac matrices,  as it was discussed in detail in Ref.~\cite{Accardi:2020iqn}. The coefficients of this expansion can only depend on the Lorentz scalars $k^2$ and $k\cdot n_{+} = k^-$. Integrating over $k^+$ and tracing over $\gamma^-$ to obtain $\alpha$, for instance, corresponds to integrating $\Xi$ over $k^2$ and tracing over $\gamma^-$. Therefore, one obtains that $\alpha$ can only be a function of $k^-$, and the same reasoning also applies to $\zeta$. On the other hand, a factor of $(k^2+\bm{k_T}^2)/2k^-$ 
appears in the analogous calculation of the trace in Eq.~\eqref{def:omega}, and $\omega$ necessarily depends on both $k^-$ and $\bm{k_T}$.

We will discuss the calculation of these coefficients in detail in Sec.~\ref{s:generic_gauge}, but it is already worthwhile comparing the free propagator of an on-shell quark of mass $m$ decomposed in light-cone coordinates, 
\begin{align}
  \slashed{k} + m = k^-\, \gamma^+ + \slashed{k}_T + m \id
    +  \frac{m^2 + \bm{k_T}^2}{2k^-}\, \gamma^- \, ,
\label{eq:quarkprop}
\end{align}
and the the quark-jet correlator~\eqref{eq:jet_decomp} in its final form, namely,
\begin{align}
\label{e:J_Dirac_explicit}
  J(k^-,\bm{k_T};n_+)
    = \frac{\theta(k^-)}{4(2\pi)^3\, k^-} \, 
    \bigg\{ k^-\, \gamma^+ + \slashed{k}_T + M_j \id + \frac{K_j^2 + \bm{k_T}^2}{2k^-} \gamma^- \bigg\} \ .
\end{align}
In the latter formula, the ``jet mass'' $M_j$ corresponds to a mass term for the hadronizing quark~\cite{Accardi:2019luo,Accardi:2020iqn}   
and the ``jet virtuality''
\begin{align}
  K_j^2 = \mu_j^2 + \tau_j^2
\end{align}
receives contributions from the invariant mass initially produced in the quark fragmentation process ($\mu_j^2$), and from the transverse broadening of the jet of particles due to final state interactions ($\tau_j^2$). 
The $\theta(k^-)$ factor in front of the curly brackets appears because the discontinuity of the jet correlator is summing over all real particles production processes in the final state~\cite{Accardi:2020iqn}, and the $M_j$ and $K_j^2$ factors are $k$-independent because of the Lorentz covariance and gauge invariance of the jet correlator $\Xi$, as we will also explicitly prove later.
Comparing the free quark propagator~\eqref{eq:quarkprop} to the jet correlator~\eqref{e:J_Dirac_explicit}, we can thus interpret the twist-3 ${\cal O}(1/k^-)$ coefficient as a gauge-invariant nonperturbative generalization of the quark's current mass,
\begin{align}
    m \leftrightsquigarrow M_j \ ,
\end{align}
with $M_j$ summing over the masses of the quark hadronizatio products \cite{Accardi:2019luo,Accardi:2020iqn}. The twist-4 ${\cal O}(1/(k^-)^2)$ coefficient can be similarly interpreted as a gauge-invariant nonperturbative generalization of the quark's invariant mass, that now includes a contribution from all hadronization products.
\begin{align}
    m^2 \leftrightsquigarrow K_j^2.    
\end{align}
It is worth emphasizing the $M_j$ and $K_j^2$ are gauge invariant quantities, although the separation of the latter into invariant mass produced during hadronization and transverse broadening of the final state depends on the choice of gauge (see Sec.~\ref{s:generic_gauge}). As with any multiparticle state, $K_j^2$ is not constrained to be equal to $M_j^2$ as it happens for a perturbative quark.

\section{Spectral analysis in the light-cone gauge}
\label{s:axial_gauges}


The jet correlator $\Xi$ is by definition gauge invariant, contrary to the quark operator $i\tilde{S}$ appearing in Eq.~\eqref{def:jetcorr} which depends on the gauge choice. This has been made explicit with the inclusion in the arguments of $i\tilde{S}$ of the 4-vector $4$-vector $v$ that defines the axial gauge condition 
\bea
v \cdot A =0 .
\label{eq:axial_gauge}
\eea
The vector $v$ can in principle be spacelike, timelike or lightlike. For our puposes, it suffices to consider the light-like axial gauge, \begin{align} 
    v^2=0 \ ,
\end{align}
in which case the most general form of the quark bilinear is given by:
\bea 
    i\widetilde{S}_{ij}(p,v) = 
    \hat{s}_3(p^2, p\cdot v)\, \slashed{p}_{ij} + 
    \sqrt{p^2}\, \hat{s}_1(p^2, p\cdot v)\, \mathbb{I}_{ij} + \hat{s}_0(p^2, p\cdot v)\, \slashed{v}_{ij}\, ,
\label{eq:qaurk_corr_decomp}
\eea
where $\hat{s}_i(p^2, p\cdot v)$ are spectral operators that are functions of all non-trivial Lorentz scalars that can be built with the 4-vectors $p$ and $v$, namely $p^2$ and $p \cdot v$. Note that we omitted structures proportional $\gamma_5$ and $[\slashed{p},\slashed{v}]$ because of parity and time-reversal invariance. 

The gauge condition \eqref{eq:axial_gauge} is in fact invariant under a $v \rightarrow \alpha v$ rescaling of the gauge vector. This implies that the operators $\hat{s}_{3,1}$ can only depend on $p^2$,
\bea
\label{e:s31_dep}
    \hat{s}_{3}(p^2, p\cdot v) = \hat{s}_{3}(p^2) 
    \, , \qquad 
    \hat{s}_{1}(p^2, p\cdot v) = \hat{s}_{1}(p^2)\, , 
\eea
while 
\bea
\label{e:s0_dep}
\hat{s}_{0}(p^2, p\cdot v) = \frac{p^2}{p\cdot v}\hat{s}_{0}(p^2) \, ,
\eea
where the $p^2$ factor is included at the numerator for dimensional purposes only. 
Thus, the quark operator in the light-cone gauge has the restricted form (omitting the Dirac indices): 
\bea
i\widetilde{S}(p, v) = \hat{s}_3(p^2)\slashed{p}+\sqrt{p^2}\hat{s}_1(p^2)\mathbb{I}+\frac{p^2}{p\cdot v} \hat{s}_0(p^2)\slashed{v}.
\label{eq:sij_lcg}
\eea
For later convenience, we also decompose each one of the $\hat{s}_{i}(p^2)$ operators into ``physical'' ($\hat\sigma_i$) and ``non-physical'' ($\hat\omega_i$) contributions,
\bea
    \hat{s}_{i}(p^2)
        =\hat{\sigma}_{i}(p^2) \, \theta(p^2)\theta(p^-) + \hat{\omega}_i(p^2)\left[1-\theta(p^2)\theta(p^-) \right],
\label{def:s:spec}
\eea
where $\theta(p^-)$ selects positive-energy states and $\theta(p^2)$ selects momenta on or inside the lightcone.

The quark propagator can be given an integral representation of the K\"{a}ll\'en-Lehmann type~\cite{Bjorken:1965zz,Weinberg:1995mt}: 
\bea
    \colav \, \langle\Omega|i\tilde S(p,v)|\Omega\rangle 
    = 
    \int \frac{d\sigma^2}{(2\pi)^4} \frac{\rho(\sigma^2)}{p^2-\sigma^2+i0} \theta(\sigma^2),
\eea
where here $\rho({\sigma}$) is a matrix in Dirac space:
\bea
\label{e:rho_sp_func}
\rho(p^2)= \rho_3(p^2) \slashed{p} +\sqrt{p^2}\rho_1(p^2)+\frac{p^2}{p\cdot v}\rho_{0}(p^2)\slashed{v}.
\eea
With the discontinuity at $p^2$ in the integrand evaluating to $(-2\pi i)\, \rho(\sigma^2)\, \delta(p^2-\sigma^2)\, \theta(p^-)$, we find that
\bea
    \mathrm{Disc} \colav \, \langle\Omega|i\tilde{S}(p,v)|\Omega\rangle 
    &=& \frac{1}{(2\pi)^3}\rho(p^2)\theta(p^2)\theta(p^-),
\label{eq:vacdisc}
\eea
and one can easily see that only the physical operators are discontinuous,
\bea
    \mathrm{Disc}\colav\,\langle\Omega|\,\hat{\sigma}_i(p^2)\,|\Omega\rangle 
    &=&  \rho_i(p^2)/(2\pi)^3,
\label{eq:vev:sigma} \\
    \mathrm{Disc}\colav\langle\Omega|\,\hat{\omega}_i(p^2)\,|\Omega\rangle 
    &=& 0 \ ,
\label{eq:vev:omega}
\eea
and by virtue of Eq.~\eqref{eq:vev:sigma} can thus be called ``spectral operators''.

Compared to expressions often found in literature, and in particular in the Refs.~\cite{Accardi:2019luo,Accardi:2020iqn} that this paper elaborates on, Eq.~\eqref{e:rho_sp_func} contains an additional spectral function $\rho_{0}(\sigma^2)$ corresponding to the $\slashed{v}$ Dirac structure. When $v$ is chosen proportional to $n_+$, this new term can bring additional contributions to the twist-$4$ coefficient $\omega(k^-,\bm{k_T})$ in Eq.~\eqref{def:omega} (see also Ref.~\cite{Accardi:2020iqn}).
This is however not the case due to a novel sum rule for the new spectral function $\rho_0$ that we derive below as a consequence of the gauge invariance of the jet correlator.

The starting point to derive the sum rule for $\rho_0$ is to consider the $J^{[\gamma^{\mp}]}$ projections of the integrated jet correlator onto $\gamma^-$ or $\gamma^{+}$, and utilize the decomposition (\ref{eq:sij_lcg}) of the quark bilinear operator:
\bea
    J^{[\gamma^{\mp}]}
    &=&  \mathrm{Disc}\int dk^{+}\int d^{4}p\,\frac{\mathrm{Tr_c}}{\mathrm{N_c}} \langle\Omega|\Big[\hat{s}_3(p^2)(p\cdot n_{\pm})+\frac{v\cdot n_{\pm}}{v\cdot p}p^2\,\hat{s}_0(p^2)\Big]\widetilde{W}(k-p;n_{+})|\Omega\rangle,
\label{eq:alphagen2} \ .
\eea
The gauge invariance of $J^{[\gamma^{\mp}]}$ then implies that the second term on the right hand side must vanish for any (light-like) vector $v$:
\bea 
    \phi^{\mp}(k^-) \equiv
    \mathrm{Disc}\int dk^{+}\int d^{4}p\,\frac{(v\cdot n_{\pm})p^2}{v\cdot p}\frac{\mathrm{Tr_c}}{\mathrm{N_c}} \langle\Omega|\hat{s}_{0}(p^2)\widetilde{W}(k-p;n_{+})|\Omega\rangle &=& 0 \, ,
\label{eq:indep_of_v}
\eea
where $\phi^\mp$ is introduced as a convenient shorthand, and the $\pm$ superscripts refer, respectively to the traces of J multiplied by $\gamma^-$ and $\gamma^+$, respectively.  
Taking  $v=n_{+}$, the function $\phi^{-}(k^-)$ vanishes identically, but for $\phi^{+}(k^-)$ we have
\bea
    \phi^{+}(k^-)
    &=& \frac{1}{2}\, \mathrm{Disc}\int dp^2\int dp^-\frac{p^2}{(p^-)^2}\frac{\mathrm{Tr_c}}{\mathrm{N_c}} \langle\Omega|\hat{s_{0}}(p^2) 
    \int \frac{d\xi^{+}}{2\pi}\, e^{i\xi^+(k^- - p^-)}\, W_{\mathrm{coll}}(\xi^+)
|\Omega\rangle, 
\label{eq:Jnminus}
\eea
where we have written the measure as $d^{4}p=dp^2\, d^2\bm{p}_T\, dp^-/2p^-$ and the integrations over $k^+$ and $\bm{p}_T$ have produced the collinear Wilson line $W_{\mathrm{coll}}(\xi^+)$.
With the specific  $v = n_+$ gauge vector choice and by virtue of our choice of staple-like Wilson lines in the
$\omega \equiv v$ direction in the definition of the jet correlator, the collinear Wilson line reduces to the identity matrix in color space: $W_{\mathrm{coll}}(\xi^+)= \id$. 
Using then the decomposition of the operator $\hat{s}_{0}$ in terms of the operators $\hat{\sigma}_0$ and $\hat{\omega}_0$ given in Eq.~\eqref{def:s:spec}, the discontinuity can be calculated using Eqs.~\eqref{eq:vev:sigma} and \eqref{eq:vev:omega}:
\bea
    \phi^{+}(k^-)
    &=& \frac{1}{2(2\pi)^3}\frac{\theta(k^-)}{(k^-)^2}\int_{0}^\infty dp^2\,p^2\, \rho_0(p^2)\, .
\eea
By virtue of Eq.~\eqref{eq:indep_of_v}, we therefore obtain a new sum rule for the $\rho_{0}(p^2)$ quark spectral function:
\bea
\label{e:sumrule_rho0}
    \int_{0}^\infty dp^2\,p^2\, \rho_0(p^2) =0 \, .
\eea
By defining the $n^{th}$-moment of a generic spectral function $\rho$ as:
\begin{equation}
\label{e:rho_moments}
\rho^{(n)} = \int_0^{+\infty} dp^2\, \big( p^2 \big)^{n/2}\, \rho(p^2) \, ,
\end{equation}
one can express the sum rule in Eq.~\eqref{e:sumrule_rho0} simply as $\rho_0^{(2)}=0$. 
Notice that this term could in principle have contributed to the twist 4 projection $J^{[\gamma^+]}$ of the integrated jet correlator in a generic axial gauge \cite{Accardi:2020iqn}. However, with the sum rule Eq.~\eqref{e:sumrule_rho0} one can explicitly see that this projection does not depend on gauge-related quantities such as $\rho_0$, as it may be expected on the basis of the gauge invariance of $J$. Sum rules for $\rho_3$ and $\rho_1$ in the light-cone $n_+ \cdot A =0$ gauge were discussed in Ref.~\cite{Accardi:2019luo,Accardi:2020iqn} and will be further examined in a generic gauge in the following section.

\section{Twist expansion in a generic gauge}
\label{s:generic_gauge}

As discussed in Ref.~\cite{Accardi:2020iqn}, the $\alpha(k^{-})$ and $\zeta(k^{-})$ coefficients of the integrated jet correlator \eqref{eq:jet_decomp} can be directly evaluated in the light-cone gauge by using the convolution representation~\eqref{def:jetcorr} and the Dirac decomposition \eqref{eq:qaurk_corr_decomp} of the quark bilinear operator. In the following, we will instead obtain expressions for these coefficients without committing to a choice of gauge and show that they are actually invariant in form and not just numerically due to gauge invariance. Moreover, the strategy employed will also allow us to complete the calculation of the twist-four projection, which is more involved and was only partially carried out in Ref.~\cite{Accardi:2020iqn} in the light-cone gauge. Here, we complete that calculation and extend it to a generic gauge.

\subsection{Twist two projection}
\label{ss:tw2}

Using the definition~\eqref{def:alpha} and after integrating over $k^{+}$ and $\vect{\bm p_{T}}$, the twist-2 $\alpha(k^{-})$ coefficient can be written as
\bea
\alpha(k^{-})= \frac{1}{2}\mathrm{Disc}{\displaystyle \int}dp^{2}\int dp^{-}\dfrac{\Tr_c}{\mathrm{N_c}}\langle\Omega |\hat{s}_{3}(p^2)\int \dfrac{d\xi^{+}}{2\pi}\, e^{i\xi^{+}(k^{-}-p^{-})}W_{\mathrm{coll}}(\xi^{+})|\Omega\rangle.
\eea
Notice that the domain of integration of $p^2$ extends over the entire real axis. The physical region, however, is constrained to $p^2>0$ and $p^- >0$ as previously discussed. Using the quark operator decomposition given in Eq.~\eqref{def:s:spec}, the only non-vanishing contribution to $\alpha(k^-)$ comes from the physical operator $\hat{\sigma}_{3}(p^2)$:
\bea
\alpha(k^{-})
&=& \frac{1}{2}\mathrm{Disc}{\displaystyle \int_{0}^{\infty}}dp^{2} \dfrac{\Tr_c}{\mathrm{\mathrm{N_c}}}\langle\Omega |\hat{\sigma}_{3}(p^2)I_{2}(k^{-})|\Omega\rangle,\nn\\
\eea
where
\bea
I_{2}(k^{-}) &=& \int_{0}^{\infty} dp^-\int \frac{d\xi^+}{2\pi}e^{i\xi^{+}(k^{-}-p^{-})}W_{\mathrm{coll}}(\xi^{+}).
\eea
%

\tocless\subsubsection{Light-cone gauge}
\label{ss:tw2_lcg}

In the light-cone gauge, the collinear Wilson line reduces to the identity in color space and the result of the integral above is straightforward,
\bea
    I_{2}(k^{-})=\int_{0}^{\infty} dp^-\;\delta(k^- -p^-)=\theta(k^-).
\label{eq:I_alpharesult}
\eea
and the twist-2 coefficient takes on a particularly simple form,
\bea
    \alpha(k^{-}) = \frac{\theta(k^-)}{2(2\pi)^3}\int_{0}^{\infty} dp^{2} \, \rho_{3}^\text{lcg} (p^2) \, ,
\label{eq:Jalpha_lcg}
\eea
where we have used the properties \eqref{eq:vev:sigma}-\eqref{eq:vev:omega} of the spectral operators, and the ``lcg'' superscript on the gauge-dependent spectral function emphasizes that, thus far, this sum rule has only been established in the light-cone gauge. 
Exploiting, furthermore, the normalization property of the $\rho_3(p^2)$ spectral function, namely $ \int_{0}^{\infty}dp^{2}\rho_3(p^2)=1$ \cite{Bjorken:1965zz,Weinberg:1995mt}, one also finds 
\bea
    \alpha(k^{-}) = \frac{\theta(k^-)}{2(2\pi)^3}.
\eea
Due to the gauge invariance of the integrated correlator $J$, and therefore of its Dirac coefficients, this result is valid in any gauge, as we will explicitly verify next.

\tocless\subsubsection{Generic gauge}
\label{ss:tw2_gg}

For the calculation in a general gauge, in which the Wilson line cannot be handled in a trivial way, it is convenient to define the dimensionless variables 
\begin{align}
    y = \xi^{+}k^{-} \qquad \text{and} \qquad \sigma = p^{-}/k^{-} \ .
\end{align}
Using these new variables and inverting the order of integration, $I_{2}(k^{-})$ can be written as
\bea
I_{2}(k^-) 
=\int \frac{dy}{2\pi}\,e^{iy}u_{k^-}(y)W_{\mathrm{coll}}(y/k^-),
\label{eq:I2}
\eea
where we have defined
\bea
u_{k^-}(y)\equiv \int_{0}^{\infty}d\sigma \,e^{-iy\, \sign{(k^-)}\, \sigma}
.
\label{eq:udef}
\eea
The $u_{k^-}(y)$ function is nothing else than a half-plane Fourier transform,
\bea
\label{e:hpFT}
\int_{0}^{\infty}d\omega\, e^{\pm ix\omega}= \pi\delta(x)\pm i\,\mathrm{P}\left(\frac{1}{x} \right),
\label{eq:pv}
\eea
where $\mathrm{P}$ denotes the Cauchy principle value \cite{Bogolyubov:1959bfo}. When using  Eq.~\eqref{eq:pv} in Eq.~\eqref{eq:I2}, one can furthermore take advantage of 
$\int_{-\infty}^{\infty}dx\, \mathrm{P}\left(\frac{1}{x}\right)e^{iy}f(x)=i\pi f(0)$ to set the Wilson line to unity, and find 
\bea
    \alpha(k^{-}) = \frac{\theta(k^-)}{2(2\pi)^3} \int_{0}^{\infty} dp^{2} \, \rho_{3} \text(p^2)
\label{eq:Jalpha}
\eea
independently of the chosen gauge. Furthermore, $\int_{0}^{\infty} dp^{2} \, \rho_{3} \text(p^2) =1$ by virtue of the canonical commutation relations \cite{Weinberg:1995mt}, and one arrives at
\bea
    \alpha(k^{-}) = \frac{\theta(k^-)}{2(2\pi)^3} \ .
\label{eq:Jalpha}
\eea

\subsection{Twist three projection}
\label{ss:tw3}
%
We can now apply the method that has just been discussed to calculate the twist-three $\zeta(k^-)$ coefficient. Using its definition~\eqref{def:zeta} and integrating over $k^+$ and $\bm{p}_{T}$ one obtains:
\bea
    \zeta(k^-)
    =
    \frac{1}{2\Lambda} \int_{0}^{\infty}dp^2\;\mathrm{Disc}\frac{\mathrm{Tr_c}}{\mathrm{N_c}}\langle\Omega|\sqrt{p^2}\,\hat{\sigma_{1}}(p^2)I_{3}(k^-)|\Omega\rangle \, ,
\label{eq:zetaJ3}
\eea
where $I_{3}(k^-)$ is defined to be:
\bea
I_{3}(k^-) 
&=& k^-\int_{0}^{\infty}\frac{dp^{-}}{p^{-}}\int\frac{d\xi^+}{2\pi}e^{i\xi^{+}(k^{-}-p^{-})}W_{\mathrm{coll}}(\xi^{+}).
\eea

\tocless\subsubsection{Light-cone gauge}
\label{ss:tw3_lcg}

As in the twist-two case, the $I_3$ integral can be directly evaluated in the light-cone gauge, leading to $I_{3}(k^-) = \theta(k^-)$.
Therefore
\bea
    \zeta(k^-)
    = \frac{\theta(k^-)}{2\Lambda(2\pi)^3}M_{j},
\label{eq:zeta:result}
\eea
where
\bea
    M_j= \int dp^2 \;\sqrt{p^2}\;\rho^{\mathrm{lcg}}_1(p^2)\, 
\label{eq:Mj_sum_rule_lcg}
\eea
is the chiral-odd jet mass~\cite{Accardi:2017pmi,Accardi:2019luo,Accardi:2020iqn} and ``lcg" emphasizes that $\rho^{\mathrm{lcg}}_1(p^2)$ is to be understood as the spectral function calculated in the light-cone gauge. While the numerical value of $M_j$ must be the same in any gauge due to the gauge invariance of the Dirac coefficients of the integrated jet correlator $J$, the same is not a priori true of the sum rule \eqref{eq:Mj_sum_rule_lcg}: in principle, in a different gauge the relation might be of the form $M_j = \rho_1^{(1)} + $``other terms''. 
Nonetheless, the invariance in form of Eq.~\eqref{eq:Mj_sum_rule_lcg} will be proven in the next subsection. 

\tocless\subsubsection{Generic gauge}
\label{ss:tw3_gg}

In a general gauge, we can once again utilize the dimensionless variables $y$ and $\sigma$ defined in Sec.~\ref{ss:tw2} and write 
\bea
    I_{3}(k^-) 
    &=&
    \sign(k^-)\int \frac{dy}{2\pi}v_{k^-}(y)\,e^{iy}\,W_{\mathrm{coll}}(y/k^-) \, ,
\label{eq:J3:dimvar}
\eea
where $v_{k^-}(y)$ is defined as
\bea
v_{k^{-}}(y)\equiv \int_{0}^{\infty}\frac{d\sigma}{\sigma}e^{-iy\sigma\sign{(k^-)}}.
\label{def:v}
\eea
Note that the $v_{k^{-}}(y)$ function is conveniently related to the $u_{k^{-}}(y)$ defined in Eq.~\eqref{eq:udef} by 
\bea
v_{k^{-}}'(y) = -i\,\sign{(k^{-})}\,u_{k^{-}}(y).
\label{eq:vprime}
\eea
In order to deal with the Wilson line, we can then write the exponential appearing in Eq.~\eqref{eq:J3:dimvar} as a derivative,  
$e^{iy}=-i\frac{\partial}{\partial y}e^{iy}$, and perform the integration over $y$ by parts. Repeating the same trick as needed, one sees that
\bea
    I_{3}(k^-) 
    = \theta(k^-)\left[\sum_{n=0}^{\infty}i^n\left(\partial_{y}^{n}W_{\mathrm{coll}}(y/k^-)\right)\right]_{y=0}
    = \theta(k^-) \left(\frac{1}{1-i\partial_y}W_{\mathrm{coll}}(y/k^-)
    \right)_{y=0}\, .
\label{e:J3Ztheta}
\eea
Substituting this result into Eq.~\eqref{eq:zetaJ3}, one obtains
\bea
    \zeta(k^-)
    =
    \frac{\theta(k^-)}{2\Lambda} \int_{0}^{\infty}dp^2\,\mathrm{Disc}\frac{\mathrm{Tr_c}}{\mathrm{N_c}}\langle\Omega|\sqrt{p^2}\,\hat{\sigma_{1}}(p^2)\left(\frac{1}{1-i\partial_y}W_{\mathrm{coll}}(y/k^-)
    \right)_{y=0}|\Omega\rangle \, .
\label{eq:zeta_as_derivative_of_W}
\eea
In the light-cone gauge, where $W_\text{coll}=\id$ we recover $I_{3}(k^-) = \theta(k^-)$ and the light-cone sum rule~\eqref{eq:Mj_sum_rule_lcg}. In a general gauge, we can expand Eq.~\eqref{eq:zeta_as_derivative_of_W} in powers of $g$ and obtain
\bea
    \zeta(k^-)
    &=& 
    \frac{\theta(k^-) }{2\Lambda} \int_{0}^{\infty}dp^2\,\mathrm{Disc}\frac{\mathrm{Tr_c}}{\mathrm{N_c}}\langle\Omega|\sqrt{p^2}\,\hat{\sigma_{1}}(p^2)\left[ 1+ \frac{g}{k^-}A^-(0) + O\Big(\frac{g}{k^-}\Big)^2  \right]|\Omega\rangle \, .
\label{eq:zetaJ3-2}
\eea
Now, compare this to the light-cone gauge result given in Eq.~\eqref{eq:zeta:result}, that depends on $k^-$ only through the $\theta(k^-)$ function. By gauge invariance of $\zeta$, this must be true also of Eq.~\eqref{eq:zetaJ3-2}. 
But nothing in the RHS can generate $k^-$-dependent terms  to cancel the explicitly written $O(g/k^-)^n$ factors when $n>0$. Therefore all the terms of $O((g/k^-)^n)$ terms must vanish with the exception of the $n=0$ term. Therefore, one obtains
\bea
    M_j = \int dp^2\, \sqrt{p^2}\, \rho_1(p^2) \, .
\eea
Recalling the definition of moments given in Eq.~\eqref{e:rho_moments}, this means that $M_j \equiv \rho_1^{(1)}$ is gauge invariant, even though the spectral function itself is not. To the best of our knowledge, this is a novel spectral function sum rule.

\subsection{Twist four projection}
\label{ss:tw4}
We now proceed to the calculation of the twist-4 $\omega$ coefficient defined in Eq.~\eqref{def:omega}. Its non-vanishing contribution in a generic gauge is given by
\bea
\omega(k^-, \bm{k_T}^2)  
&=& \left(\frac{k^-}{\Lambda} \right)^2\mathrm{Disc}\int dk^{+}\int d^{4}p\,\frac{\mathrm{Tr_{c}}}{\mathrm{N_c}} \langle\Omega|\hat{\sigma}_3(p^2)\frac{p^2+\bm{p_T}^2}{2p^-}\widetilde{W}(k-p;n_{+})|\Omega\rangle,
\label{eq:omega}
\eea
where we have written 
$p\cdot n_-=(p^2+\bm{p_T}^2)/2p^-$.  The integration over $k^+$ simply sets the conjugate space coordinate to zero in the partial Fourier transform of the Wilson line reducing this to $W_\text{TMD}$. In the term term proportional to $p^2$ the integration over the transverse momentum then acts directly over $W_\text{TMD}$, further reducing the Wilson line to its collinear $W_\text{coll}$ form and simplifying the calculation.
In a general gauge, where $W_\text{coll} \neq \id$, the calculation is more complicated but can be performed by following the strategy employed for the twist-2 and twist-3 coefficients. However, the integration of the second term in the integrand of Eq.~\eqref{eq:omega} is substantially more involved due to the presence of $\bm{p_T}^2$ which prevents one from directly integrating the Wilson line over its transverse momentum. We thus write Eq.~\eqref{eq:omega}, in general, as a sum of two terms and will address these in turn:
\bea
\omega(k^-, \bm{k_T}^2)  
&=& \omega_\ell(k^-)+\omega_t(k^-, \bm{k_T}^2),
\label{eq:omegasum}
\eea
where $\omega_\ell$ and $\omega_t$ are, respectively, the transverse-momentum-independent (``longitudinal'') and transverse-momentum-dependent (``transverse'') components of the twist-4 coefficient:
\begin{align}
    \omega_\ell(k^-)  
        &= \frac{1}{(2\Lambda)^2}\int_{0}^{\infty} dp^2 p^2\,\mathrm{Disc}\frac{\mathrm{Tr_{c}}}{\mathrm{N_c}}\langle\Omega|\hat{\sigma}_3(p^2) I_\ell(k^-)|\Omega\rangle, 
        \label{eq:omega1} \\
    \omega_t(k^-, \bm{k_T}^2)  
        &= \frac{1}{(2\Lambda)^2}\int_{0}^{\infty} dp^2 \,\mathrm{Disc}\frac{\mathrm{Tr_{c}}}{\mathrm{N_c}}\langle\Omega|\hat{\sigma}_3(p^2)I_t(k^-, \bm{k_T})|\Omega\rangle, 
        \label{eq:omega2}
\end{align}
where
\begin{align}
    I_\ell(k^-)
        & = \int_0^{\infty}dp^-\left(\frac{k^-}{p^-}\right)^2\int \frac{d\xi^+}{2\pi}e^{i\xi^+(k^--p^-)}W_{\mathrm{coll}}(\xi^+), 
    \label{eq:J4} \\
    I_t(k^-, \bm{k_T})
        &= \int_{0}^\infty dp^- \left(\frac{k-}{p^-}\right)^2\int d^2 p_T\int \frac{d^2{\bf \xi}_{T}}{(2\pi)^2}\,\vect{p_T}^2\,e^{i\vect{\xi_{T}}\cdot(\bm{k_T}-\vect{p_T})} \widetilde{W}_{\mathrm{TMD}}( k^--p^-, \bm{\xi_T}).
    \label{def:Jomega}
\end{align}

\tocless\subsubsection{Light-cone gauge}
\label{ss:tw4_lcg}

In the light-cone gauge, the calculation of the longitudinal term is straightforward, and one finds 
\begin{align}
    I_\ell^{\mathrm{lcg}}(k^-)
      & = \theta(k^-)
      \label{eq:Jl:lcg}
      \\
    \omega_\ell^{\mathrm{lcg}}(k^-) 
        &
        =\frac{\theta(k^-)}{(2\Lambda)^2(2\pi)^{3}}(\mu_{j}^2)^{\mathrm{lcg}},
\label{eq:omega1:lcg}
\end{align}
where
\bea
    (\mu_{j}^2)^{\mathrm{lcg}} 
    \equiv \int_{0}^{\infty}dp^2\;p^2\;\rho_3^{\mathrm{lcg}}(p^2)
\eea
can be interpreted as the average invariant mass squared of the particles produced by the quark fragmentation~\cite{Accardi:2020iqn}. 

The calculation of the transverse $I_t$ operator in Eq.~\eqref{def:Jomega} is more involved. First, we can eliminate the $\bm{p_T}^2$ term by using
\bea
\bm{p_T}^2\,e^{i\vect{\xi_{T}} \cdot (\bm{k_T}-\vect{ p_{T}})} = 
\left[(i\vect{ \partial_{T}})^2+\bm{k_T}^2+2\,i\,\bm{k_T}\cdot\vect{ \partial_{T}}\right]
e^{i\vect{\xi_{T}} \cdot (\bm{k_T}-\vect{p_T})} \ ,
\label{eq:ident}
\eea
where we shortened $\bm{\partial_T} = \partial/\partial\bm{\xi_T}$.
Then, the integration over the term proportional to $\bm{k_T}^2$ presents no difficulty, and proceeds as for $I_{\ell}$ in Eq.~(\ref{eq:Jl:lcg}). In the remaining terms, integration by parts over $d^2\bm{\xi_T}$ produces derivatives of the $W_\text{TMD}$ Wilson line with respect to  the transverse space coordinate. In the light-cone gauge, $W_\text{TMD}$ reduces to a transverse gauge link at light-cone infinity, which can also be set to unity by imposing advanced boundary conditions on the transverse gauge fields, namely $\vect{A_T}(\infty^+)=0$~\cite{Belitsky:2002sm,Gao:2013rba}. 
The transverse derivatives are then identically equal to zero and we obtain a simple enough final result:
\bea
I_t^{\mathrm{lcg}}(k^-, \bm{k_T})
&=& \theta(k^-)\, \bm{k_T}^2,
\nn\\
\omega_t^{\mathrm{lcg}}(k^-, \bm{k_T})
&=& \frac{\theta(k^-)}{(2\Lambda)^2(2\pi)^3}\, \bm{k_T}^2.
\eea
Together with Eq.~\eqref{eq:omega1:lcg} this leads to the full result for the twist 4 projection in the light-cone gauge,
\bea
   \omega^{\mathrm{lcg}}(k^-, \bm{k_T}^2)  
   &=& \frac{\theta(k^-)}{(2\Lambda)^2(2\pi)^3}\left[(\mu_j^2)^{\mathrm{lcg}}+\bm{k_T}^2\right] .
\label{eq:omega_lcg}
\eea
This has a simple interpretation: the invariant mass $m_q^2$ of the (single particle) quark is replaced by the non-perturbative invariant mass $\mu_j^2$ of the particles produced in the quark's fragmentation process, while the $k_T^2$ contribution is of purely kinematic origin. However, unlike the twist-3 $\zeta$ coefficient, this simple interpretation is not gauge invariant as we will discuss next.

\tocless\subsubsection{Generic gauge}
\label{ss:tw4_gg}

In a generic gauge, the longitudinal $I_\ell$ and $\omega_\ell$ can be obtained by following the strategy deployed for the calculation of the twist-2 and twist-3 projections. One obtains
\bea
    I_{\ell}(k^-) 
    &=& 
    \theta(k^-) \left[\frac{1}{(1-i\partial_y)^2}W_{\mathrm{coll}}(y/k^-)
    \right]_{y=0},
    \nn\\
    \omega_\ell(k^-)
    &=&  \frac{\theta(k^-)}{(2\Lambda)^2(2\pi)^3} 
    \int dp^2 \,p^2\, \rho_3(p^2) +
    O\left(\frac{g}{k^-}\right). 
\label{eq:omega:l}
\eea

To calculate the transverse part one needs to evaluate $I_t$ and $\omega_t$. The starting point is to use Eq.~\eqref{eq:ident} to remove the $\bm{p_T}^2$ dependence in the integrand of Eq.~\eqref{def:Jomega} with transverse derivatives. After integration by parts, we are left with first and second transverse derivatives of the Wilson line. By symmetry arguments, the term proportional to the first derivative can be shown to vanish upon integration.
However the second derivative of the Wilson line, that vanished in the light-cone gauge, will give a nontrivial contribution to the twist-4 coefficient. Upon recursively using 
$e^{iy}=-i\frac{\partial}{\partial y}e^{iy}$ as for the twist-3 calculation, we find that
\bea
    I_t(k^-, \bm{k_T}) 
        &=& \theta(k^-)\left[\left( \bm{k_T}^2 -\vect{\partial_{T}}^2 \right)\frac{1}{(1-i\partial_y)^2}W_{\mathrm{TMD}}(y/k^-,\bm{\xi_T})
        \right]_{y,\bm{\xi_T}=0},
    \nn\\
    \omega_t(k^-, \bm{k_T}^2)  
        &=& \frac{\theta(k^-)}{(2\Lambda)^2}\int_{0}^{\infty} dp^2 \,\mathrm{Disc}\frac{\mathrm{Tr_{c}}}{\mathrm{N_c}}\langle\Omega|\hat{\sigma}_3(p^2)\left[\frac{\left( \bm{k_T}^2 -\vect{\partial_{T}}^2 \right)}{(1-i\partial_y)^2}W_{\mathrm{TMD}}(y/k^-,\bm{\xi_T})
        \right]_{y,\bm{\xi_T}=0}|\Omega\rangle\, .
    \label{eq:omega:t}     
\eea
Expanding the latter in powers of $g/k^-$ we find 
\bea
    \omega_t(k^-, \bm{k_T}^2) 
    =
    \frac{\theta(k^-)}{(2\Lambda)^2(2\pi)^3}\bm{k_T}^2 
         + \frac{\theta(k^-)}{(2\Lambda)^2}\int_{0}^{\infty} dp^2 \,\mathrm{Disc}\frac{\mathrm{Tr_{c}}}{\mathrm{N_c}}\langle\Omega|\hat{\sigma}_3(p^2)\bm{J_T}|\Omega\rangle + O\left(\frac{g}{k^-}\right),
         \label{eq:omega:t2} 
\eea
where
\bea
    \bm{J_T}
    =\left[ -\vect{\partial_{T}}^2  W_{\mathrm{TMD}}(y/k^-,\bm{\xi_T})\right]_{y, \bm{\xi_T}=0}.
    \label{def:Jperp}
\eea
The detailed evaluation of the above derivative can be found in Appendix~\ref{a:W_second}.   
The result is
\bea
\bm{J_T}
&=& ig \,\vect{D_{T}} \vect{A}_{T}(0)
\label{eq:Jperp:result} 
\\  
&+&ig\left.\left[\int_{0}^{\infty^+} ds^+ \vect{D_{T}}\bigg(\mathcal{U}_{n_{+}}[0^-,0^+,\vect{\xi}_{T};0^-,s^+,\vect{\xi}_{T}]\bm{G}^{\bm T -}(0^-,s^+,\vect{\xi}_{T}) \, 
\mathcal{U}_{n^+}[0^-,s^+,\bm{\xi_T};0^-,\infty^+,\vect{\xi}_{T}]\bigg)\right]\right|_{\bm{\xi_T}=0} \, ,
\nonumber
\eea
where $\vect{D_{T}}= \vect{\partial_T}+ ig \vect{A_{T}}$ is the transverse covariant derivative and $\bm{G^{T -}}$ is the field strength tensor, and repeated transverse T indexes are contracted in the 2D transverse Euclidean space.

In the light-cone gauge, $\bm{J_T}$ and therefore $\omega_t$ vanish as already implied by Eq.~\eqref{eq:omega_lcg}. It is interesting to see how this explicitly works out in the general expression we just derived. Indeed, in the light-cone gauge $\bm{G}^{\bm T -}=-\partial^{-} {\bf A}_{T}$ and the integral in Eq.~\eqref{eq:Jperp:result} evaluates to $\vect{D_{T}} \vect{A}_{T}(\infty) - \vect{D_{T}} \vect{A}_{T}(0)$. The term evaluated at the origin then cancels the first term in Eq.~\eqref{eq:Jperp:result}, and the term evaluated at infinity vanishes by imposing advanced boundary conditions, meaning that indeed $\bm{J_T}^\text{lcg} = 0$. 
It is also interesting to notice that this cancellation is actually quite general within the class of local contour gauges~\cite{Ivanov:1985np,Anikin:2021oht,Anikin:2021osx}, to which the light-cone gauge supplemented with boundary conditions pertain. The first term in Eq.~\eqref{eq:Jperp:result} is actually a residual gauge footprint left behind by not completely fixing the gauge. 
The contour gauge, in contrast, is by construction free of gauge redundancies and, applied to the staple-like Wilson line that appears throughout this work, implies that $W_{\mathrm{TMD}}=\id$, leading to $\bm{J_T}=0$ in any contour gauge.

The cancellation of the first term itself (if not of the whole $\bm{J_T}$) happens in fact in any gauge. This can be seen by perturbatively expanding the $\mathcal{U}$ contributing to the line integral of Eq.~\eqref{eq:Jperp:result} in powers of $g$, and explicitly integrating the leading term. 
The $\bm{J_T}$ operator is then proportional to $g^2$ and can be interpreted as describing two gluon rescatterings along the Wilson line from $0^+$ to $\infty^+$.

We can now quote the result for the full $\omega$ projection in a generic gauge. First, we plug Eq.~\eqref{eq:Jperp:result} back in Eq.~\eqref{eq:omega:t2} and add the longitudinal part of the $\omega$ projection. Then, we realize that by gauge symmetry all terms of order $g/k^-$ must be zero, because in the light cone gauge $\omega$ is independent of $k^-$ apart from the theta function, but the lack of any dependence on $k^-$ in the integrand in the second term in Eq.~\eqref{eq:omega:l} prevents any cancellation of the $k^-$ in the higher order terms. Finally, we find that
\bea
    \omega(k^-,\bm{k_T}) = 
    \frac{\theta(k^-)}{(2\Lambda)^2(2\pi)^3}
    \left( 
    K_j^2 + \bm{k_T}^2
    \right),
\eea
where the jet virtuality
\bea
    K_j^2 = \mu_j^2  + \tau_j^2
\label{eq:Kj_decomposition}
\eea
gets contributions from two distinct sources. The first one is the invariant mass of the quark hadronization products,
\begin{align} 
    \mu_j^2 = \int_0^\infty dp^2 \,p^2\, \rho_3(p^2) \ ,
\end{align}
which is calculated as the second moment of the chiral-even spectral function in the considered gauge and, at variance with $M_j$, turns out to not be gauge invariant as we shall see in a moment. The second source is 
the transverse momentum broadening of the final state particles,
\begin{align}
    \tau_j^2 
        &= (2\pi)^3 \int_{0}^{\infty} dp^2 \,\mathrm{Disc}\frac{\mathrm{Tr_{c}}}{\mathrm{N_c}}\langle\Omega|\hat{\sigma}_3(p^2)ig \,\vect{D_T}\left( \vect{A_T}(\bm{\xi_T})+\vect{\mathcal{Z}_T}(\bm{\xi_T})\right)_{\bm{\xi_T}=0}|\Omega\rangle,
\label{eq:tauj_general_gauge}
\end{align}
with
\bea
    \vect{\mathcal{Z}_T}(\bm{\xi_T})= ig
    \int_{0}^{\infty^+} ds^+ \mathcal{U}_{n_{+}}[0^-,0^+,\vect{\xi_{T}};0^-,s^+,\vect{\xi_{T}}]\bm{G}^{\bm T -} (0^-,s^+,\vect{\xi_{T}}) \, 
    \mathcal{U}_{n^+}[0^-,s^+,\vect{\xi_{T}};0^-,\infty^+,\vect{\xi_{T}}]|\Omega\rangle \, .
\label{eq:final-state-interactions}
\eea
In interpreting this expression, one should keep in mind that $\bm{A_T}$ appears in Eq.~\eqref{eq:tauj_general_gauge} only as a residue  of incomplete gauge-fixing procedures, as remarked earlier, so that final state rescattering effects are effectively contained in $\bm{D_T}\bm{\mathcal{Z}_T}$ only. 

Because of the gauge invariance of the integrated jet correlator, the $\omega$ coefficient is also gauge invariant. However, the decomposition \eqref{eq:Kj_decomposition} in invariant mass of the hadronization products and transverse momentum broadening in the final state is not. 
For example, in the the light-cone gauge with appropriate boundary conditions no rescattering seems to occur because  $\tau_j^\text{lcg}=0$. On should however notice that, although final state interactions seemingly disappear with the appropriate choice of gauge and boundary conditions, their effect is subsumed into $(\mu_j^2)^\text{lcg}$ -- in fact, embedded inside the chiral-even quark spectral function $\rho_3^\text{lcg}$. 
In a specific calculation, one would have to balance the simplicity of a vanishing line integral in Eq.~\eqref{eq:final-state-interactions}, with the added complexity found when solving the fully dressed quark propagator to obtain the quark spectral functions. Indeed, the boundary condition for the gauge fields at light-cone infinity implies a specific choice of pole prescription for the gluon propagator appearing in the calculation, which renders it more complex than in, say, the Landau gauge (where, however, one would also need to calculate the $\tau_j$ term). \\

\section{Conclusions}
\label{s:conclusions}

This paper has completed the spectral analysis of the gauge invariant quark propagator initiated in Ref.~\cite{Accardi:2019luo,Accardi:2020iqn} in the light cone gauge, and now extended to a generic gauge. In particular, the coefficients of the Dirac decomposition of the gauge-invariant quark propagator~\eqref{e:invariant_quark_correlator_W} have been identified with specific moments of the chiral even and chiral odd quark spectral functions. The gauge invariance of the formulas so obtained has been investigated, and novel spectral function sum rules derived.

In Sec.~\ref{ss:tw2} and~\ref{ss:tw3} it is shown that the twist-2 and twist-3 coefficients $\alpha(k^-)$ and $\zeta(k^-)$ are proportional \textit{in any gauge} to the zeroth moment $\rho_3^{(0)}$ and first moment $\rho_1^{(1)}$ of the chiral even and odd spectral functions, respectively, even though the spectral functions themselves depend on the chosen gauge. The former result squares with the well known $\int dp^2\rho_3(p^2) = 1$ sum rule, which depends only on the properties of equal time commutators of the quark fields, and is indeed independent of the chosen gauge. The latter result is particularly interesting because it shows that the jet mass $M_j = \int dp^2 \sqrt{p^2} \rho_1(p^2)$ (a color-screened gauge-invariant dressed mass for the quark) can always be interpreted as the dynamically generated mass in the quark fragmentation process, and provides one with the means of actually \textit{measuring} this, for example, in inclusive DIS \cite{Accardi:2008ne} or in semi-inclusive electron-positron annihilation \cite{Accardi:2022oog}. On the theoretical side, the gauge invariance of $\rho_1^{(1)}$ is a feature that should be confirmed by independent calculations.

In Sec.~\ref{ss:tw4} the calculation of the twist-4 term previously analyzed in Ref.~\cite{Accardi:2020iqn} has been completed. The decomposition of this coefficient function in terms of the invariant mass of the hadronizing quark and the effect of final-state interactions is elucidated. This decomposition is not gauge invariant, but the sum of the two contributions is, with final state interactions disappearing in the light cone gauge.

Finally, the presence of an additional quark spectral function in axial gauges has been discussed in Sec.~\ref{s:axial_gauges}. With considerations based on the Dirac structure of the propagator and on the gauge invariance of the jet correlator, it has been shown that the second moment of this spectral function, $\rho_0^{(2)}$, vanishes identically. We look forward to independent investigations of this new sum rule as well as numerical confirmations of the gauge invariance of $\rho_1^{(1)}$, for example, with a calculation of the light cone gauge quark spectral functions that is yet to appear in the literature.

\begin{acknowledgments}
We thank A.~Bashir, J.~Goity, G.~Krein, A.~Pilloni, M.~Sievert for helpful and inspiring discussions. 
This work was supported in part by the  U.S. Department of Energy (DOE) contract DE-AC05-06OR23177, under which Jefferson Science Associates LLC manages and operates Jefferson Lab. 
AA also acknowledges support from DOE contract DE-SC0008791.
AS received support from the European Commission through the Marie Sk\l{}odowska-Curie Action SQuHadron (grant agreement ID: 795475). 
\end{acknowledgments}

\begin{appendix}

\section{Light-cone basis}
\label{a:conv}

In a given reference frame, we collect the space-time components of a four-vector $a^\mu$ inside round parentheses, $a^\mu=(a^0,a^1,a^2,a^3)$, with $a^0$ the time coordinate. 
We define the light-cone $\pm$ components of the $a$ vector as
\begin{align}
  a^{\pm} = \frac{1}{\sqrt{2}} (a^0 \pm a^3) 
\end{align}
and collect these inside square brackets: $a^\mu = [a^-,a^+,\vect{a}_T]$, with $\vect{a}_T=(a^1,a^2)$ being the 2-dimensional components in transverse space. 
The transverse component of a vector $a^\mu$ is $a_T^\mu = [0,0,\vect{a}_T]$, such that $a_T^2 = -\vect{a}_T^2$. Namely, the norm of $\vect{a}_T$ is taken according to the Euclidean metric $\delta_T^{ij}=\text{diag}(1,1)$, whereas the norm of $a_T$ is calculated using the Minkowski metric $g^{\mu\nu}=\text{diag}(1,-1,-1,-1)$. 
Note that, in this paper, we consider highly boosted quarks and hadrons with dominant momentum component along the negative 3-axis, namely along the negative light-cone direction. Hence, we grouped the light-cone components inside the square parenthesis starting with the minus component.

The light-cone basis vectors are defined as:
\begin{equation}
\label{e:def_np_nm}
n_{\pm} = \frac{1}{\sqrt{2}} (1,0,0,\pm 1) \ \ , 
\end{equation}
such that $n_+^2 = n_-^2 =0$, $n_+^\mu n_{-\mu} = 1$, and $a^\pm = a^\mu {n_\mp}_\mu$. The transverse basis vectors in 2D Euclidean space are $\bm{n_1} = (1,0)$ and $\bm{n_2} = (0,1)$, corresponding to $n_1 = [0,0,\bm{n_1}]$ and $n_2 = [0,0,\bm{n_2}]$ in Minkowski space.

Following Ref.~\cite{Bacchetta:2006tn,Mulders:2016pln}, the transverse projector, $g_T^{\mu\nu}$, and the transverse anti-symmetric tensor, $\epsilon_T^{\mu\nu}$, are defined as:
\begin{align}
\label{e:gT_def}
g_T^{\mu\nu} & \equiv g^{\mu\nu} - n_+^{ \{ \mu} n_-^{\nu \} } \\ 
\label{e:epsT_def}
\epsilon_T^{\mu\nu} & \equiv \epsilon^{\mu \nu \rho \sigma} {n_-}_\rho {n_+}_\sigma  \equiv \epsilon^{\mu \nu + -} \ ,
\end{align}
where $g^{\mu\nu}$ is the Minkowski metric, $\epsilon^{\mu\nu\rho\sigma}$ is the totally anti-symmetric Levi-Civita tensor (with $\epsilon^{0123}=1$).
Note that $g_T^{\mu\nu} a_\nu = a_T^\mu$ projects a four-vector onto its transverse component, and $\epsilon_T^{\mu\nu} a_\nu = \epsilon_T^{\mu\nu} a_{T\nu}$ rotates that component by 90 degrees in the transverse plane.

\section{Second derivative of a Wilson line}
\label{a:W_second}

Here we outline the derivation of Eq.~\eqref{def:Jperp}, which involves second derivatives of the staple-like Wilson line with respect to the transverse direction. The derivative with respect to any point in the path can be cast in the following  form~\cite{Lorce:2012ce,Belitsky:2005qn}:
\begin{align}
    \frac{\partial}{\partial z^{\mu}}W(y,x) 
    & = 
    igW(x,s)A_{\alpha}(s)\frac{\partial s^{\alpha}}{\partial z^{\mu}}W(s,y) |_{s=y}^{s=x}+ ig{\displaystyle \int_{y}^{x} }W(x,s)G_{\alpha\beta}(s)W(s,y)\dfrac{\partial s^{\alpha}}{\partial z^{\mu}}ds^{\beta},
\label{eq:WLder3}
\end{align}
where $y$ and $x$ are the starting and ending points of the Wilson line, respectively. 
Applying Eq.~\eqref{eq:WLder3} to the staple Wilson line given in Eq.~\eqref{def:WTMD} we get that the derivative of the Wilson line with respect to $\bm{\xi_T}$ is given by:
\begin{align}
    \frac{\partial}{\partial \bm{\xi_T}}W_{\mathrm{TMD}}(\xi^+,\bm{\xi_T}) 
    &= 
    ig \vect{A_T}(\xi^+,\bm{\xi_T})W_{\mathrm{TMD}}(\xi^+,\bm{\xi_T}) + \vect{\mathcal{Z}_T}(\xi^+,\bm{\xi_T})
    \label{eq:Wder}
\end{align}
where $\bm{\mathcal{Z}_T}(\xi^+,\bm{\xi_T})$ is the line integral piece of the Wilson line derivative:
\begin{align}
    \vect{\mathcal{Z}_T}(\xi^+,\bm{\xi_T}) 
    & = 
    -ig\int_{\infty^+}^{\xi^+} ds^+ \mathcal{U}_{n_{+}}[0^-,\xi^+,\bm{\xi_T};0^-,s^+,\bm{\xi_T}]\bm{G}^{\bm{T} -}(0^-,s^+,\bm{\xi_T}) \, 
    \mathcal{U}_{n^+}[0^-,s^+,\bm{\xi_T};0^-,\infty^+,\bm{\xi_T}]
\end{align}
Notice the negative sign in front of the integral, which is due to the derivative acting on a Wilson path oriented in the negative (ligtcone plus) direction. Taking another derivative with respect to $\bm{\xi_T}$, we have: 
\begin{align}
    \frac{\partial^2}{\partial  \vect{\xi_T}^2}W_{\mathrm{TMD}}(\xi^+,\bm{\xi_T})
    &=  ig \, \frac{\partial}{\partial\vect{\xi_T}} \vect{A_T}(\xi^+,\bm{\xi_T})W_{\mathrm{TMD}}(\xi^+,\bm{\xi_T})
    + ig \, \vect{A_T}(\xi^+,\bm{\xi_T}) \frac{\partial}{\partial\vect{\xi_T}} W_{\mathrm{TMD}}(\xi^+,\bm{\xi_T}) 
    + \frac{\partial}{\partial\vect{\xi_T}} \vect{\mathcal{Z}_T}(\xi^+,\bm{\xi_T})
    \nn\\
    &= ig \vect{D_{T}}\left[ \vect{A_{T}}(\xi^+,\bm{\xi_T})W_{\mathrm{TMD}}(\xi^+,\bm{\xi_T})   +\vect{\mathcal{Z}_T}(\xi^+,\bm{\xi_T})\right] \, ,
\label{eq:WTMD_2der}
\end{align}
where repeated transverse indexes are contracted in the 2D Euclidean transverse space.
In the first line, we used Eq.~\eqref{eq:Wder} for the derivative of the Wilson line appearing in the second term.  This gives rise to two terms, one that is combined with the first term to form a covariant derivative and the other that is also combined into a covariant derivative with the last term. Writing explicitly the action of the covariant derivative on the second term, one obtains:
\begin{align}
    \frac{\partial^2}{\partial\vect{\xi_T}^2}W_{\mathrm{TMD}}(\xi^+,\bm{\xi_T})
    \nn\\
    &\hspace{-2.65cm} = 
    ig  \vect{D_{T}} \vect{A_{T}}(\xi^+,\bm{\xi_T})W_{\mathrm{TMD}}(\xi^+,\bm{\xi_T})
    + ig\int_{\xi^+}^{\infty^+} \!\!\!\!\!\!ds^+\, \mathcal{U}_{n_{+}}[\xi^+,\bm{\xi_T};s^+,\bm{\xi_T}]\vect{D_{T}}\bm{G}^{\bm{T} -}(s^+,\bm{\xi_T}) \,
    \mathcal{U}_{n^+}[s^+,\bm{\xi_T};\infty^+,\bm{\xi_T}]
    \nn\\
    &\hspace{-2.65cm} -(ig)^2\int_{\xi^+}^{\infty^+} \!\!\!\!\!\!ds^+ \!\! \int_{s^+}^{\infty^+} \!\!\!\!\!\!dr^+\, \mathcal{U}_{n_{+}}[\xi^+,\bm{\xi_T};s^+,\bm{\xi_T}]\bm{G}^{\bm{T} -}(s^+,\bm{\xi_T})\mathcal{U}_{n^+}[s^+,\bm{\xi_T};r^+,\bm{\xi_T}]\bm{G}^{\bm{T} -}(r^+,\bm{\xi_T}) \, 
    \mathcal{U}_{n^+}[r^+,\bm{\xi_T};\infty^+,\bm{\xi_T}] \, .
\end{align}
As noticed in the main text, the first term appears as an artifact of an incomplete gauge fixing procedure, which cancels with boundary contributions from the line integrals. [This can be explicitly seen by expanding the $U$ Wilson lines in the first line integral in powers of the $g$ coupling constant, and integrating the leading $ig\int ds^+ \vect{D_{T}}\bm{G}^{\bm{T}-}(s^+,\bm\xi_{\bm T})$ term.]. The second derivative of the Wilson line can then be interpreted as describing two-gluon rescattering along the straight path from $\xi^+$ to $\infty^+$ at a fixed transverse coordinate $\bm{\xi_T}$. \\

\end{appendix}

\bibliographystyle{JHEP}  
\bibliography{main.bib}

\end{document}